\documentclass[10pt]{article}
\usepackage{amsmath}
\usepackage{amssymb, amscd, amsthm}
\usepackage[all]{xy}
\usepackage[dvips]{graphicx}
\usepackage{verbatim}
\usepackage[perpage,symbol*]{footmisc}
\usepackage{cite}
\usepackage[colorlinks, linkcolor=red,anchorcolor=blue, citecolor=green]{hyperref}
\usepackage{longtable}
\newcommand{\tabincell}[2]{\begin{tabular}{@{}#1@{}}#2\end{tabular}}
\usepackage{multirow}

\begin{document}
\title{\bf New Classes of Entanglement-assisted Quantum MDS Codes}
\author{Renjie Jin\footnote{jinrenjie@mails.ccnu.edu.cn (R.~Jin)}, Derong Xie\footnote{derongxie@yahoo.com (D.~Xie)} and Jinquan Luo\footnote{Corresponding author,luojinquan@mail.ccnu.edu.cn (J.~Luo)}\footnote{ The authors are with School of Mathematics and Statistics $\&$ Hubei Key Laboratory of Mathematical Sciences, Central China Normal University, Wuhan China, 430079.}}
\date{}
\maketitle
\medskip

{\bf Abstract} \, In this paper, two new classes of entanglement-assisted quantum MDS codes\;(EAQMDS codes for short) with length $n$ being a factor
of $q^2\pm 1$ are presented via cyclic codes over finite fields of odd characteristic. Among our constructions, there are serval EAQMDS codes with new parameters which have never been reported. Moreover, some of them have much larger minimum distance than known results.

{\bf Key words} \, MDS code, EAQEC code, EAQMDS code, Cyclic code

\section{Introduction }

\noindent Quantum information can protect messages between sender and receiver avoiding decoherence by encoding it into quantum
error-correcting codes. Entanglement-assisted quantum error-correcting codes\;(EAQEC codes for short)\;are crucial to quantum information theory\;$(\text{see} [\ref{1},\ref{2},\ref{3},\ref{4},\ref{10}])$.
Recently, construction of good quantum codes via classical codes is a hot topic for quantum information and quantum computing $(\text{see} [\ref{13},\ref{17},\ref{19},\ref{21},\ref{22}]).$
EAQEC codes use preexisting entanglement between the sender and the receiver to improve information rate.
Many researchers have been devoted to obtaining EAQEC codes via classical liner codes, such as negacyclic codes and generalized Reed-Solomon codes.
It has been shown that EAQEC codes have some advantages over standard stabilizer
codes. For example, only a dual-containing classical linear quaternary code can be
transformed into a standard stabilizer code, but any classical linear quaternary code
can be transformed into an EAQEC code.
Some of them can be summarized as follows.

In $[\ref{16},\ref{20}]$, some new EAQEC codes with good parameters via cyclic and constacyclic codes are constructed. In $[\ref{5}]$, new decomposition of negacyclic codes are proposed, by which four new classes of EAQEC codes have
been constructed. In $[\ref{6}]$, Fan et al. constructed some classes of EAQMDS codes based on classical maximum distance separable\;(MDS for short)\; codes by exploiting one or more pre-shared maximally
entangled states. In $[\ref{19}]$, Qian and Zhang constructed some new classes of MDS linear complementary dual\;(LCD) codes with respect to Hermitian inner product. As applications, they have constructed new families of EAQMDS codes. In $[\ref{8}]$, Guenda et al. showed that the number of shared pairs required to construct an EAQEC code is related to the hull of classical codes. Using this fact, they put forward new methods to construct EAQEC codes requiring desirable amounts of entanglements.
The $EA$-$Singleton\; bound$ for an $[[n,k,d;c]]_{q}$ EAQEC code is $$2(d-1)\leq n-k+c.$$

A $q$-ary EAQEC code attaining this bound is said to be an EAQMDS code. In this paper, we will construct EAQMDS codes via cyclic codes by improving the method introduced in  $[\ref{20}]$.

Our main contribution is the construction of EAQMDS codes with parameters\\\\

(1) $\left[\left[n,n-4qm+4m^2+3,2m(q-1);(2m-1)^{2}\right]\right]_{q}$  where $1\leq m \leq \lfloor\frac{q+1}{4t}\rfloor,\; n=\frac{q^2-1}{t} \;\text{and}\; t\mid q^2-1.$\;(Theorem 3.1)\\\\

(2)$\left[\left[n,n-4mq+4q+4m^2-8m+3,2(m-1)q+2;4(m-1)^2+1\right]\right]_{q}$ where $2\leq m \leq\lfloor\frac{q+1}{4t}\rfloor,\;n=\frac{q^2+1}{t}\;\text{and} \;t\mid q^2+1.$\;(Theorem 3.2)\\

 We will present some known results of EAQMDS codes, which are depicted in Table 1.
\begin{center}
\begin{longtable}{|l|l|l|}  
\caption{Some known constructions on EAQMDS codes with parameters $[[n,k,d;c]]_q$} ( $q$ is an odd prime power ) \\ \hline
Parameters & Constraints & Reference\\ \hline
$[[q^{2}+1,q^{2}-4(m-1)(q-m-1),$
$2(m-1)q+2;4(m-1)^{2}+1]]_{q}$
  &  $q\geq 5$ and $2\leq m\leq\frac{q-1}{2}$   & [\ref{20}] \\  \hline

$[[q^{2}+1,q^{2}-2q-4m+5,$
$2m+q+1;4]]_{q}$ & \tabincell{c}{$q\geq 5$, $q=4t+1$ where $t$ \\is an integer and $2\leq m\leq\frac{q-1}{2}$} & [\ref{5}]\\  \hline

$[[\frac{q^{2}+1}{2},\frac{q^{2}+1}{2}-2q-4m+5,$
$2m+q+1;5]]_{q}$ & $q>7$ and $2\leq m\leq\frac{q-1}{2}$ &   [\ref{5}] \\  \hline

$[[\frac{q^{2}-1}{5},\frac{q^{2}-5q-20m+4}{5},$
$\frac{4m+q+5}{2};4]]_{q}$ & \tabincell{c}{$q=20t+3$ or $q=20t+7$ \\and $t\leq m \leq\frac{q-3}{4}$} &   [\ref{C}]\\  \hline

$[[\frac{q^{2}+1}{10},\frac{q^{2}+1}{10}-2d+3,$
$d;1]]_{q}$ & \tabincell{c}{$q=10m+3$ and $d$ is even,\\ $2\leq d \leq 6m+2$} &  [\ref{L}] \\  \hline

$[[\frac{q^{2}+1}{10},\frac{q^{2}+1}{10}-2d+3,$
$d;1]]_{q}$ & \tabincell{c}{$q=10m+7$ and $d$ is even,\\ $2\leq d \leq 6m+4$} &   [\ref{L}] \\  \hline

$[[\frac{q^{2}-1}{h},\frac{q^{2}-1}{h}-2d+3,$
$d;1]]_{q}$ & \tabincell{c}{$h\in \{3,5,7\}$ is a factor\\ of $q+1$ and $d$ is even,\\ $\frac{q+1}{h}\leq d\leq\frac{(q+1)(h+3)}{2h}-1$}  &   [\ref{L}] \\  \hline

$[[\frac{q^{2}-1}{t},\frac{q^{2}-1}{t}-4qm+4m^2+3,$
$2m(q-1);(2m-1)^{2}]]_{q}$ & \tabincell{c}{$q\geq 3$, $t\mid q^2-1$,\\ $1\leq m\leq\lfloor\frac{q+1}{4t}\rfloor$} &  Theorem 3.1 \\  \hline

\tabincell{c}{$[[\frac{q^{2}+1}{t},\frac{q^{2}+1}{t}-4qm+4q+4m^2-8m+3,$\\
$2q(m-1)+2;4(m-1)^2+1]]_{q}$}  & \tabincell{c}{$q\geq 7$, $t\mid q^2+1$,\\ $2\leq m\leq\lfloor\frac{q+1}{4t}\rfloor$}  & Theorem 3.2 \\  \hline
\end{longtable}
 \end{center}

This paper is organized as follows. In section 2, we will introduce some basic acknowledge and useful results on cyclic codes and EAQEC codes.
In section 3, we will persent our main results on the constructions of new EAQMDS codes. In section 4, we will make a conclusion.\\\\

\section{Preliminaries}
\subsection{Cyclic Code}
\noindent In this section, we introduce some basic notations and useful results on linear codes and cyclic codes.
Let $\mathbb{F}_{q}$ be the $finite \;field$ of cardinality $q$, where $q$ is an odd prime power. An $[n,k,d]_{q}$ code is an
$\mathbb{F}_{q}$-linear subspace of $\mathbb{F}^{n}_{q}$ with dimension $k$ and minimum distance $d.$
The $ Singleton\; bound$ states that
$$ d\leq n-k+1.$$

The code attaining the Singleton bound is called $MDS$ code. When $d=n-k$, the code is called almost MDS. Let $\mathbf{u}=(u_{0},\cdots,u_{n-1})$ and $\mathbf{v}=(v_{0},\cdots,v_{n-1})$ be two vectors in $ \mathbb{F}_{q^2}^{n}.$ The $ Hermitian\;
inner\; product$ is defined by $$ \langle \mathbf{u},\mathbf{v} \rangle_{H}=u_{0}v_{0}^{q}+u_{1}v_{1}^{q}+\cdots+u_{n-1}v_{n-1}^{q}.$$
The $Hermitian\; dual$ of an $\mathbb{F}_{q^2}$-linear code $\mathcal{C}$ of length $n$ is defined as $$\mathcal{C}^{\perp_{ H}}=\{\mathbf{u}\in F_{q^2}^{n} \mid \langle \mathbf{u},\mathbf{v} \rangle_{H}=0 \;\text{for\,all}\,\mathbf{v}\in \mathcal{C}\}.$$
The code $\mathcal{C}$ is $Hermitian\; self$-$orthogonal$ if $\mathcal{C} \subseteq \mathcal{C}^{\perp_{ H}}$, and is  $Hermitian\; self$-$dual$ if $\mathcal{C}= \mathcal{C}^{\perp_{ H}}$.\\

For $ \gcd(n,q)=1$, the $q^2$-$cyclotomic\,\; coset$ of $i$ modulo $n$ is defined by\\
$$ C_{i}=\{iq^{2j}(\text{mod}\, \,n )\mid j\in \mathbb{Z\}}.$$
A linear code $\mathcal {C}$ over $\mathbb{F}_{q^2}$ is a $cyclic \,\;code$ if for every codeword $(c_{1},c_{2},\cdots,c_{n})\in\mathcal {C}$, its cyclic shift $(c_{n},c_{1},\cdots,c_{n-1})$ is also a codeword in $\mathcal {C}$.
The cyclic code $\mathcal {C}$ can be generated by a polynomial $g(x),$ where $g(x)\mid x^n-1.$
The $defining\,\; set$ of $\mathcal {C}$ is given by $T=\{0\leq i \leq n-1\mid g(\alpha^{i})=0\},$
where $\alpha$ is an $n$-th root of unity in some extension field of $\mathbb{F}_{q^{2}} .$ It is easy to see that
the defining set $T$ is a union of some $q^2$-cyclotomic cosets. Then the following property is given in $[\ref{14},\ref{15}]$.\\\\

{\bf Proposition 2.1}\quad (BCH bound)\;Let $\delta$ be a positive integer with $2\leq \delta \leq n.$ Assume that
$\mathcal{C}$ is a cyclic code of length $n$ with defining set $T$. If $T$ contains $\delta-1$ consecutive elements $ \alpha^{b},\alpha^{b+1},\cdots,\alpha^{b+\delta-2}$, then minimum distance of $\mathcal{C}$ is at least $\delta$.

\subsection{EAQEC code}

\noindent In this section, we introduce some notations and useful results on EAQEC codes. For more details, see $[\ref{6},\ref{7},\ref{9},\ref{11},\ref{18}].$

A $q$-ary $[[n,k,d;c]]_{q}$ $EAQEC$ code $\mathcal{C}$ has length $n$  can encode $k$ logical qubits with minimum distance $d$. Here,
$c$ is the copies of maximally entangled Bell states. The code $\mathcal{C}$ can correct up to at least $[\frac{d-1}{2}]$ quantum errors.

Recently, researchers proved that EAQEC code can be constructed  from any classical linear code over $\mathbb{F}_{q} $. The remaining problem is to calculate $c.$ In $[\ref{12}]$, a new approach to determine $c$ is proposed by Li et al. \\\\

{\bf Proposition 2.2} $([\ref{12}])$ \quad Let $\mathcal{C}$  be an $[n,k,d]_{q^2}$ cyclic code with defining set $T$. Assume the decomposition of
$T$ is $T=T_{ss}\bigcup T_{as}$, where $T_{ss}=-qT\bigcap T$ and $T_{as}=T\backslash T_{ss}$.

(1) Let $\mathcal{C}_{1}$ and $\mathcal{C}_{2}$ be cyclic codes with defining set $T_{ss}$ and $T_{as}$ respectively. Then $\mathcal{C}_{1}^{\perp _{H}}\bigcap \mathcal{C}_{1}=\{0\}$ and $\mathcal{C}_{2}^{\perp_{ H}}\subseteq \mathcal{C}_{2}.$

(2) There exists an $ [[n,n-2|T|+|T_{ss}|,d;|T_{ss}|]]_{q} $ EAQEC code.

\section{New EAQMDS Codes }

\subsection{Length $n\mid q^2-1$}
 \noindent In this subsection, we apply cyclic codes of length $n\mid q^2-1$ to construct a new family of EAQMDS codes with length $n\mid q^2-1$, where $q\geq 3$ is an odd prime power. Firstly, the $q^2$-ary cyclotomic coset modulo $n$ are singletons:\;$C_{i}=\{ i\;(\text{mod} \;n)\}$.
The following two lemmas will be used in our constructions.\\\\

{\bf Lemma 3.1} \quad Let $q$ be an odd prime power, $t\mid q^2-1$ and $n=\frac{q^2-1}{t}.$ Then\\
 $$-qC_{aq+b}=C_{-bq-a}$$
where $-\lfloor\frac{q-1}{2t}\rfloor\leq a,b\leq\lfloor\frac{q-1}{2t}\rfloor,$ with $(a,b)\neq (-\lfloor\frac{q-1}{2t}\rfloor,-\lfloor\frac{q-1}{2t}\rfloor).$\\\\

{\bf Proof}\quad Note that $|aq+b|\leq \lfloor\frac{q^2-1}{2t}\rfloor$  and $aq+b\neq -\lfloor\frac{q^2-1}{2t}\rfloor.$ Hence, all of $aq+b$ are distinct. A straightforward calculation shows\\
\begin{align*}
-q(aq+b)&=-aq^2-bq\\
&=-a(q^2-1)-a-bq\\
&\equiv-bq-a \pmod{n}, \end{align*}
which implies $-qC_{aq+b}=C_{-bq-a}$. \quad\quad \quad \quad \quad \quad \quad \quad \quad \quad \quad\quad \quad\quad \quad\quad \quad\quad \quad \quad \quad \quad \quad \quad \quad \quad \quad  $\Box$\\

{\bf Lemma 3.2}\quad Let $m$ be a positive integer with $1\leq m \leq\lfloor\frac{q+1}{4t}\rfloor$ and $n\mid q^2-1$ where $q$ is an odd prime power. Decompose  $$T_{0}=\bigcup\limits_{-m\leq i \leq m-1,i\neq0} A_{i}$$  where
\begin{equation*}
A_{i}=\left\{
\begin{array}{lll}

\{-mq+j \mid m+1\leq j \leq q-m\}, & & {i=-m};\\\\

\{iq+j  \mid m\leq j \leq q-m\}, & & {-m+1 \leq i \leq m-2,\; i\neq 0};\\\\

\{mq+j \mid -q+m\leq j \leq -m-1\}, & & {i=m-1}.\end{array} \right.\end{equation*}
Then  $-qT_{0}\bigcap T_{0}=\emptyset\;.$\\

{\bf Proof}\quad To show  $-qT_{0}\bigcap T_{0}=\emptyset,\;$ it suffices to prove the following cases.\\
\begin{itemize}
\item {\bf Case 1:}\quad $-qA_i\bigcap A_{i'}=\emptyset$. On the contrary, suppose there exist $j,j^{'}$ such that
\begin{align*}
-q(iq+j)\;
  &=-iq^2-jq\;(\text{mod}\;n)\\
&=-jq-i\;(\text{mod}\;n). \end{align*}
It is easy to verify that
$$i^{'}=-j\; \text{and} \; j^{'}=-i.$$
It follows that $ -(q-m)\leq i^{'}\leq -m\; \text{and} \; -(m-2)\leq j^{'}\leq m-1$. Since $-m+1\leq i\leq m-2$ and $m\leq j\leq q-m$,
 which leads to a contradiction.\\
 Hence $ -qA_{i} \bigcap A_{i^{'}}=\emptyset. $ \\

\item {\bf Case 2}:\quad $-qA_i\bigcap A_{-m}=\emptyset$ and $-qA_i\bigcap A_{m}=\emptyset$.
On the contrary, suppose $-qA_i\bigcap A_{-m}\neq\emptyset$, then there exist $j,j^{'}$ such that
 \begin{align*}
-q(iq+j)&=-jq-i\;(\text{mod}\;n)\\
&=-mq+j^{'}\;(\text{mod}\;n).\\
 \end{align*}
where $m\leq j\leq q-m \;\text{and}\; -q+m\leq j^{'}\leq -m-1$. As a consequence, $-(m-2)\leq -i\leq m-1$.  which is a contradiction.

Assume $-qA_i\bigcap A_{m}\neq\emptyset$, then there exist $j,j^{''}$ such that
 \begin{align*}
-q(iq+j)&=-jq-i\;(\text{mod}\;n)\\
&=mq+j^{''}\;(\text{mod}\;n).\\
 \end{align*}
Where $m\leq j\leq q-m \;\text{and}\; m+1\leq j^{''}\leq q-m$. As a consequence, $-(q-m)\leq -j\leq -m$.  It is easy to verify $-qA_i\bigcap A_{m}=\emptyset$.\\

\item {\bf Case 3}:\quad $-qA_{-m}\bigcap A_{m}=\emptyset$ and $-qA_m\bigcap A_{-m}=\emptyset$.\\
On the contrary, suppose $-qA_{-m}\bigcap A_{m}\neq\emptyset$, then there exist $j,l^{'}$ such that
\begin{align*}
  -q(-mq+j)
  &=m(q^2-1)-jq+m\;(\text{mod}\;n)\\
  &=-jq+m\;(\text{mod}\;n)\\
  &=mq+l^{'}\;(\text{mod}\;n).\\\end{align*}
Where $m+1\leq j\leq q-m$ and $-(q-m)\leq l^{'}\leq -(m+1)$, which is a contradiction.\\
 Hence $-qA_{-m}\bigcap A_{m}=\emptyset$.
Assume  $-qA_m\bigcap A_{-m}\leq\emptyset$, then there exist $j,l^{''}$ such that
\begin{align*}
  -q(mq+j)
  &=-m(q^2-1)-jq-m\;(\text{mod}\;n)\\
  &=-jq-m\;(\text{mod}\;n)\\
  &=-mq+l^{''}\;(\text{mod}\;n).\\\end{align*}
Where $-(q-m)\leq j\leq -(m+1)$ and $m+1\leq l^{''}\leq q-m$.  Which is a contradiction.\\
 Hence $-qA_{-m}\bigcap A_{m}=\emptyset$.

\item The remaining cases $-qA_{-m}\bigcap A_{-m}=\emptyset$ and $-qA_m\bigcap A_{m}=\emptyset$ can be proved in a similar way and we omit the details.
\end{itemize}
Hence we have $-qT_{0}\bigcap T_{0}=\emptyset\;,$ which completes the proof. \quad \quad \quad \quad \quad \quad \quad \quad \quad \quad \quad \quad \quad \quad \quad  $\Box$\\

 Now we will construct new EAQMDS codes with length $n \mid q^2-1$.\,\\\\

 {\bf Theorem 3.1}\quad Let $q$ be an odd prime power, $t\mid q^2-1$ and $n=\frac{q^2-1}{t}$.\,For any $1\leq m \leq \lfloor\frac{q+1}{4t}\rfloor$ there exists EAQMDS code with parameters
$$[[n,n-4qm+4m^2+3,2m(q-1);(2m-1)^{2}]]_{q}.\;$$\\

{\bf Proof}\quad For $1\leq m \leq \lfloor\frac{q+1}{4t}\rfloor,\;$ assume that $\mathcal{C}$ is a cyclic code of length $n\mid q^2-1$ with defining set\\
\begin{equation*}
  T=\bigcup_{i=-mq+m+1}^{mq-m-1}C_{i}
\end{equation*}

It is easy to check that $\mathcal{C}$ has $2qm-2m-1$ consecutive roots.\, Then by Proposition 2.\,1,\;the minimum distance of $\mathcal{C}$ is at least $2qm-2m.\,$ It follows that
$\mathcal{C}$ is a cyclic code with parameters $[n,\;n-2qm+2m+1,\;2qm-2m]_{q^2}.\,$
In the following, we show that $c=|T_{ss}|=(2m-1)^{2}$.\,\\
Denote by
\begin{equation*}
  T_{0}=\bigcup_{i=-m,i\neq 0}^{m-1}A_{i}.
\end{equation*}
According to Lemma 3.\,2,\; $-qT_{0}\bigcap T_{0}=\emptyset.\;$  By Lemma 3.\,1,\; \\
$$ -q C_{1}= C_{-q},\;-q C_{q+1}= C_{-q-1},\;\cdots,\;-q C_{(m-1)q-1}= C_{q-(m-1)},\;$$
$$-q C_{2}= C_{-2q},\;-q C_{q+2}= C_{-2q-1},\;\cdots,\;-q C_{(m-1)q-2}= C_{2q-(m-1)},\;$$
$$ \vdots $$
$$-q C_{m-1}= C_{-(m-1)q},\;-q C_{q+(m-1)}= C_{-(m-1)q-1}\cdots,\;-q C_{(m-1)q-(m-1)}= C_{(m-1)q-(m-1)}.\,$$
Obviously $-q C_{0}= C_{0}$  and it suffices to prove  \\
$$T_{ss}=-qT\;\bigcap\;T=\bigcup _{-(m-1)\leq a , b \leq m-1}  C_{aq+b}.$$\\
For $ C_{aq+b}\subseteq T_{ss}$, there exist $a^{'},b^{'}$ such that $-qC_{aq+b}=C_{a^{'}q+b^{'}}$.
The equality $$-q(aq+b)\equiv a^{'}q+b^{'}\;(\text{mod}\;n)$$   implies $$a^{'}=-b,b^{'}=-a.$$

Then by Lemma 3.1, $-qT_{ss}=T_{ss}$ and by Lemma 3.2  $-qT_{0}\bigcap T_{0}=\emptyset.\;$ By the standard counting arguments, $\left|T_{ss}\right|=(2m-1)^{2}.\,$\\
By Proposition 2.\,2, the EAQEC code with defining set $T$ has parameters\\
$$[[n,n-4qm+4m^2+3,2m(q-1);(2m-1)^{2}]]_{q}.\,$$
Also this code reaches the $EA$-$Singleton$ bound,\;
$$n-k+c+2=4qm-4m=2d.\,$$
Hence,\; the EAQMDS code with desired parameters are constructed. \quad \quad \quad\quad \quad  \quad \quad\quad \quad\quad \quad \quad  $\Box$\\

{\bf Example 3.\,1}\quad We list some new parameters of EAQMDS codes  of Theorems 3.1 in Table 2.
\begin{center}
\begin{longtable}{lll}  
\caption{Some new parameters of EAQMDS codes } \\  \hline

Parameters & $t$ & $m$\\ \hline
$[[280,171,56;1]]_{29}$  & 3  & 1   \\  \hline

$[[280,67,112;9]]_{29}$  & 3  & 2  \\   \hline

$[[560,403,80;1]]_{41}$  & 3  & 1  \\   \hline

$[[560,251,160;9]]_{41}$  & 3  & 2  \\   \hline

$[[368,187,92;1]]_{47}$  & 6  & 1  \\    \hline

$[[368,9,184;9]]_{47}$  & 6  &  2    \\   \hline

\end{longtable}
 \end{center}

\subsection{Length $n|q^2+1$}

\noindent In this subsection,\; we try to construct some new EAQMDS codes with length $n\;|\;q^2+1$. In this section, we assume $n$ is even. Denote by $t=\frac{q^2+1}{n}$(it is an integer). The $q^2$-ary cyclotomic coset modulo $n$ are\\
$$ C_{0}=\{0\},\; C_{1}=\{1,\;n-1\},\; C_{2}=\{2,\;n-2\},\;\cdots,\; C_{\frac{n}{2}}=\left\{\frac{n}{2}\right\}.\,$$
The following two lemmas are similar to Lemmas 3.\,1 and 3.\,2.\,\\\\

{\bf Lemma 3.\,3} \quad Let $n=\frac{q^2+1}{t}$. Then\\
 $$-q C_{cq+d}= C_{dq-c}\;,$$\\
 where $1\leq c \leq\lfloor\frac{q-1}{2t}\rfloor$ and $0\leq d \leq \lfloor\frac{q-1}{2t}\rfloor $.\,\\

{\bf Proof}\quad Note that $ C_{cq+d}=\{cq+d,\;-(cq+d)\}\; \text{with}\;\;cq+d\leq \lfloor\frac{q^2-1}{2t}\rfloor$ for $1\leq c \leq\lfloor\frac{q-1}{2t}\rfloor$ and $0\leq d \leq \lfloor\frac{q-1}{2t}\rfloor $.\,A straightforward calculation shows
\begin{align*}
-q\cdot(-(cq+d))&=cq^2+dq\\
&=c(q^2+1)-c+dq\\
&\equiv dq-c\pmod{n}\\\end{align*}
which implies $-q C_{cq+d}= C_{dq-c}$.\,\quad \quad \quad \quad\quad\quad \quad \quad \quad \quad \quad \quad \quad \quad\quad \quad \quad\quad \quad \quad \quad \quad \quad\quad \quad\quad \quad \quad  $\Box$\\

{\bf Lemma 3.\,4}\quad Let $n=\frac{q^2+1}{t}$,\;notations as in Lemma 3.3. For $2\leq m \leq\lfloor\frac{q+1}{4t}\rfloor$,\, let
$$T_{1}=\bigcup_{0\leq c\leq m-2,\;m \leq d\leq \lfloor\frac{q-1}{2t}\rfloor} C_{cq+d}\bigcup_{1\leq e\leq m-1\leq f\leq\lfloor\frac{q-1}{2t}\rfloor} C_{eq-f}.\,$$
 Then $-qT_{1}\bigcap T_{1}=\emptyset$.\,\\

 {\bf Proof}\quad By Lemma 3.\,3,\;
$$-qT_{1}=\bigcup_{0\leq c\leq m-2,\;m \leq d\leq \lfloor\frac{q-1}{2t}\rfloor} C_{dq-c}\bigcup_{1\leq e\leq m-1\leq f\leq\lfloor\frac{q-1}{2t}\rfloor} C_{fq+e}.\,$$
When $m \leq d\leq \lfloor\frac{q-1}{2t}\rfloor$ and $\;0\leq c\leq m-2$,\;

$$cq+d\leq (m-2)q+\lfloor\frac{q-1}{2t}\rfloor\quad\text{and}\quad mq-m+2\leq dq-c.\,$$
Similarly
$$eq-f\leq (m-1)q+1-m\quad\text{and}\quad (m-1)q+1\leq fq+e.\,$$
It is easy to verify
$$cq+d\leq dq-c,\;cq+d\leq fq+e,\;eq-f\leq dq-c,\;eq-f\leq fq+e.\,$$
Therefore $-qT_{1}\bigcap T_{1}=\emptyset$.\,\quad \quad \quad \quad\quad\quad \quad \quad \quad\quad \quad\quad \quad \quad \quad \quad\quad \quad \quad\quad \quad \quad \quad \quad \quad\quad \quad\quad \quad \quad  $\Box$\\

{\bf Theorem 3.\,2}\quad Let $n=\frac{q^2+1}{t}$. For any $2\leq m \leq\lfloor\frac{q+1}{4t}\rfloor,$ there exists an EAQMDS code with parameters\\
$$\left[\left[n,n-4mq+4q+4m^2-8m+3,2(m-1)q+2;4(m-1)^2+1\right]\right]_{q}$$\,\\

{\bf Proof}\quad For $2\leq m \leq\lfloor\frac{q+1}{4t}\rfloor$,\, suppose  $\mathcal{C}$ is a cyclic code with length $n=\frac{q^2+1}{t}$
and defining set $T= C_{0}\bigcup C_{1}\bigcup C_{2}\bigcup\cdots\bigcup C_{(m-1)q}.\;$ It is easy to see that $\mathcal{C}$ has
$2(m-1)q+1$ consecutive roots.\, Then by Proposition 2.\,1,\;the minimum distance of $\mathcal{C}$ is at least $2(m-1)q+2.\,$ It follows that
$\mathcal{C}$ is a cyclic code with parameters $[n,\;n-2(m-1)q-1,\;2(m-1)q+1]_{q^2}.\,$\\
In the following, we show that $\left|T_{ss}\right|=4(m-1)^{2}+1.\,$
Let
$$T_{1}=\bigcup_{0\leq c\leq m-2,\;m \leq d\leq \lfloor\frac{q-1}{2t}\rfloor} C_{cq+d}\bigcup_{1\leq e\leq m-1\leq f\leq\lfloor\frac{q-1}{2t}\rfloor} C_{eq-f}.\,$$
According to Lemma 3.\,3,\; \\
$$ -q C_{1}= C_{q},\;-q C_{q+1}= C_{q-1},\;\cdots,\; -qC_{(m-2)q+1}= C_{q-(m-2)},\;$$
$$\quad -q C_{2}= C_{2q},\;-q C_{q+2}= C_{2q-1},\;\cdots,\;-q C_{(m-2)q+2}= C_{2q-(m-2)},\;$$
$$ \vdots $$
$$\quad\quad -q C_{m-1}= C_{(m-1)q},\;-q C_{q+(m-1)}= C_{(m-1)q-1}\cdots,\;-q C_{(m-2)q+(m-1)}= C_{(m-1)q-(m-2)}.\,$$
 Obviously $-q C_{0}= C_{0}.\,$ It is easy to verify that
 $ T_{ss}=T\backslash T_{1}.\,$ Then $-qT_{ss}=T_{ss}$ and $-qT_{1}\bigcap T_{1}=\emptyset.$ It is easy to see $|T_{ss}|=4(m-1)^2+1.$
By Proposition 2.\,2 the EAQEC code with defining set $T$ has parameters
$$\left[\left[n,n-4mq+4q+4m^2-8m+3,2(m-1)q+2;4(m-1)^2+1\right]\right]_{q}$$ Also this code reaches the $EA$-$Singleton$ bound,
$$n-k+c+2=4(m-1)q+4=2d.\,$$
Hence,\; the code $\mathcal{C}$ is an EAQMDS code.\,\quad\quad \quad \quad \quad \quad \quad \quad \quad \quad \quad \quad\quad \quad \quad \quad \quad \quad\quad \quad\quad \quad \quad  $\Box$\\

 {\bf Example 3.\,2}\quad We list some new parameters of EAQMDS codes  of Theorems 3.2 in Table 3.\\\\

\begin{center}
\begin{longtable}{lll}  
\caption{Some new parameters of EAQMDS codes } \\  \hline

Parameters & $t$ & $m$\\ \hline
$[[370,201,88;5]]_{43}$  & 5  & 2   \\  \hline

$[[466,113,180;9]]_{89}$  & 17  & 2  \\   \hline

$[[898,633,136;5]]_{67}$  & 5  & 2  \\   \hline

$[[898,377,270;17]]_{67}$  & 5  & 3  \\   \hline

\end{longtable}
 \end{center}

{\bf Remark 3.1} \quad From Examples 3.1 and 3.2, we can conclude that the required parameters of the EAQMDS codes constructed in Theorems 3.1 and 3.2 are more flexible than
all codes listed in Table 1, since our results covers almost all possible factors of $q^2\pm 1$. What's more, the parameters $c$ in our codes is unfixed. Compared to [\ref{20}],
we introduced a more general method (see Lemma 3.2 ) to find $T_{as}$ and $T_{ss}$, then we solved the case  length $n\mid q^2-1$.
Since the four parameters in our codes are flexible, it is easier to obtain a large number of EAQMDS codes from our constructions than those listed in Table 1.
 In [\ref{LI}], the authors constructed EAQMDS codes with length $n=\frac{q-1}{a}(q+1)$. Since the length of their codes satisfies $n\mid q-1$,
our length can be a factor of $q+1$. Then we can construct some EAQMDS codes with new parameters that have never been reported. Besides, employing the EAQMDS codes
obtained by Theorems 3.1 and 3.2, we can obtain EAQMDS codes with length different from $q + 1$ and the required parameters can take all or almost all possible values.
Some of them are listed in Table 2.

\section{Conclusion}

\quad In this paper,\;we construct two new classes of EAQMDS codes with length $n|q^2-1$ and $n|q^2+1$ via classical cyclic codes.\;Our
codes have more flexible parameters than known EAQEC codes. It may be possible to apply our methods to construct new EAQMDS codes via classical
 linear codes, such as generalized Reed-Solomon codes or constacyclic codes.\\

\end{document}